\documentclass{article}

\usepackage[english]{babel}

\usepackage[letterpaper,top=2cm,bottom=2cm,left=3cm,right=3cm,marginparwidth=1.75cm]{geometry}

\usepackage{amsmath,amsfonts,amssymb}
\usepackage{algpseudocode}
\usepackage{graphicx}
\usepackage[colorlinks=true, allcolors=blue]{hyperref}
\usepackage{subcaption}
\usepackage{authblk}

\title{Data-driven stochastic spectral modeling for coarsening of the two-dimensional Euler equations on the sphere}
\author[1]{Sagy R. Ephrati \footnote{Corresponding author (s.r.ephrati@utwente.nl)}}
\author[1]{Paolo Cifani}
\author[2]{Milo Viviani}
\author[1,3]{Bernard J. Geurts}
\affil[1]{Mathematics of Multiscale Modeling and Simulation, Faculty EEMCS, University of Twente, 7500 AE Enschede,
The Netherlands}
\affil[2]{Scuola Normale Superiore di Pisa, Pisa, Italy}
\affil[3]{Multiscale Energy Physics, CCER, Faculty Applied Physics, Eindhoven University of Technology,
5600 MB Eindhoven, The Netherlands}

\usepackage{ulem}

\usepackage{color}
\newcommand{\sre}[1]{\textcolor{black}{{#1}}}

\newcommand{\add}[1]{\textcolor{blue}{#1}}

\begin{document}
\maketitle


\begin{abstract}
A resolution-independent data-driven stochastic parametrization method for subgrid-scale processes in coarsened fluid descriptions is proposed. The method enables the inclusion of high-fidelity data into the coarsened flow model, thereby enabling accurate simulations also with the coarser representation. The small-scale parametrization is introduced at the level of the Fourier coefficients of the coarsened numerical solution. It is designed to reproduce the kinetic energy spectra observed in high-fidelity data of the same system. The approach is based on a control feedback term reminiscent of continuous data assimilation. The method relies solely on the availability of high-fidelity data from a statistically steady state. No assumptions are made regarding the adopted discretization method or the selected coarser resolution. The performance of the method is assessed for the two-dimensional Euler equations on the sphere. Applying the method at two significantly coarser resolutions yields good results for the mean and variance of the Fourier coefficients. Stable and accurate large-scale dynamics can be simulated over long integration times.
\end{abstract}

\section{Introduction}
Two-dimensional incompressible hydrodynamics models are fundamental for studying physical phenomena in atmospheric and oceanic flows. Typical examples include the two-dimensional Euler equations, quasi-geostrophic equations, and (rotating) shallow water equations. A characteristic feature of these flows is the formation of both large vorticity structures through the inverse energy cascade and small-scale vorticity filaments through the enstrophy cascade \cite{zeitlin2018geophysical}. In realistic conditions, the energy spectrum extends over several orders of magnitude, making it computationally infeasible to fully resolve all scales that are present in the flow. Simplifications are required, either by reducing the complexity of the underlying mathematical model \cite{luesinkthesis} or by reducing the spatial or temporal resolution with which the dynamics are resolved \cite{bos2010}. In this paper, we will focus on high-fidelity coarsening of the two-dimensional Euler equations on the sphere by applying an \textit{online/offline} approach to obtain accurate coarse-grained numerical solutions of statistically steady states. In particular, explicit information on well-resolved dynamics is obtained from high-resolution simulations in the offline phase, which is applied in an online control feedback model for accurate coarse-grained simulations.  \\
There is considerable interest in achieving accurate numerical solutions of fluid flows at reduced computational costs \cite{geurts2022book}. This forms the main challenge of Large Eddy Simulation (LES), which aims to provide skillful large-scale predictions of complex flows by numerically solving spatially filtered momentum equations. Often, a model term is included to compensate for unresolved dynamics due to coarsening to retain a sufficiently detailed description of turbulent flows at high Reynolds numbers \cite{geurts2003elements, sagaut2006large,geurtsholm2002,piomelli2015,rouhi2016}. The growing availability of computational resources has facilitated the use of high-resolution direct numerical simulations (DNS) as a source of data from which coarse-grid fluid models may be derived. Data-driven LES methods have successfully been developed in recent years, for example, by using neural networks to compute a variable eddy viscosity \cite{beck2019deep} to approximate a reference kinetic energy spectrum \cite{kurz2023deep} or to model subgrid scale-scale forces \cite{xie2020modeling}. Alternatively, approaches based on interpolation of small high-resolution patches of the spatial domain \cite{cao2016multiscale, bunder2021large} and data-driven residual modeling via global basis functions \cite{ephrati2022computational} have also shown computational efficiency and accuracy in coarse-grained numerical solutions.\\
Data assimilation provides an alternative method to achieving accurate coarse-grained results by combining predictions with real-time observations. In continuous data assimilation (CDA), observational data is incorporated into the prediction while the numerical model is being integrated in time \cite{altaf2017downscaling, charney1969use, daley1993atmospheric}. Specifically, the difference between the numerical prediction and the corresponding observation determines a nudging term that is added to the governing equations. Studies on nudging of dissipative fluid models have shown that a range of nudging strengths may be chosen that all yield an accurate coarse-grained representation of the true solution \cite{azouani2014continuous, altaf2017downscaling}. Adaptive nudging strengths based on energy balance have also been proposed \cite{zerfas2019continuous} resulting in faster convergence towards the reference compared to a simulation that exploits a constant nudging strength. Since these models rely on observational data to achieve high-fidelity coarsened solutions, the uncertainty originating from measurement errors has to be taken into account, as well as possible accumulation of discretization errors \cite{geurtsbos2005}.\\
Models of geophysical fluid flows often employ stochasticity as a means to model uncertainty inherent to flows \cite{palmer2019stochastic}. Uncertainty arises predominantly from differences in initial conditions, errors in measurements, and model incompleteness. 
Low-dimensional models describing qualitative features of geophysical fluid flows often serve as a test bed for stochastic forcing. For example, stochastic forcing based on subgrid data in the two-scale Lorenz '96 system resulted in improved forecasting skill compared to deterministic parametrizations \cite{arnold2013stochastic}. Ultimately, the exact way in which stochasticity is included in numerical simulations remains a modeling choice and may lead to qualitatively different effects on the dynamics \cite{geurts2020lyapunov}. These approaches have also been applied successfully to more complete geophysical models. Examples include the modeling of uncertainty through Casimir-preserving stochastic forcing for the two-dimensional Euler equations \cite{cotter2019numerically, ephrati2023data, cifani2023sparse} and energy-preserving stochastic forcing in the quasi-geostrophic equations \cite{resseguier2020data}. An alternative approach is based on statistics of subgrid data that lead to a stochastic forcing and eddy viscosity, which has been applied to the barotropic vorticity equation on the sphere \cite{frederiksen2006dynamical}. This approach was found to accurately model uncertainty and produce energy spectra on coarse computational grids that closely match reference high-fidelity simulations at much higher resolutions.\\
In this paper, we propose an \textit{online} data-driven standalone stochastic model for coarse numerical simulations of statistically steady states of the two-dimensional Euler equations on the sphere. Data of a statistical equilibrium is extracted from an \textit{offline} high-resolution precursor simulation in the form of statistics of coefficients of spherical harmonic modes and is included as a stochastic forcing term closely following the formulation of the continuous-time limit of the 3D-Var algorithm \cite{courtier1998ecmwf} as presented in \cite{blomker2013accuracy}. Similar to data-driven LES, a modeling term is added to the coarsened numerical simulation based on these {\it{a priori}} collected data. This term models the unresolved interactions between the modes as a linear stochastic process for each spherical harmonic coefficient separately and is designed to reproduce the energy spectrum of the high-resolution simulation. Like CDA, the model term is included as a feedback control term. This term nudges the coarse grid solution towards a known reference solution, chosen here as the statistically steady state. We opt for the nudging strength to be equal to the inverse of the characteristic time scale of the corresponding spherical harmonic mode. This choice has the benefit that it mimics the measured temporal correlation. The nudging procedure is performed via a prediction-correction scheme in which we first fully complete a time integration step involving all true fluxes and subsequently we apply the nudge as a correction to the predicted solution. This results in straightforward implementation in existing computational methods and leads to a numerical scheme of the same form as the diagonal Fourier domain Kalman filter \cite{harlim2008filtering, majda2012filtering} with prescribed gain. Striking features of the high-fidelity reference solution were captured in the coarser model using this stochastic model.\\
The paper is structured as follows. The two-dimensional Euler equations and the adopted numerical method are introduced in Section \ref{sec:governingeqs}. In Section \ref{sec:model} we describe the model and focus in particular on how the model parameters are specified. In Section \ref{sec:results}, we define the reference solution and apply the model at two coarse resolutions. The results are assessed qualitatively and by means of statistics of Fourier coefficients. Subsequently, we show that the model is capable of reproducing large-scale vortex dynamics of the reference solution. Section \ref{sec:conclusion} concludes the paper and suggests directions for further research. 

\section{Governing equations and numerical methods}\label{sec:governingeqs}
The model that will be studied in this work is given by the two-dimensional Euler equations on the unit sphere $\mathbb{S}^2$. These equations arise as the two-dimensional Navier-Stokes equations in the inviscid limit and describe vortex dynamics \cite{zeitlin2018geophysical}. The dynamics are given in streamfunction-vorticity formulation by \begin{equation}
    \begin{aligned}
        \dot{\omega} & = \left\{\psi, \omega \right\}, \\
        \Delta\psi &= \omega.
    \end{aligned}
    \label{eq:Euler}
\end{equation}
Here $\omega$ is the vorticity, $\psi$ is the streamfunction, and $\left\{\cdot, \cdot \right\}$ is the Poisson bracket. 
The vorticity and the streamfunction are related via the Laplace operator $\Delta$. The vorticity relates to the fluid velocity $\mathbf{v}$ via $\omega = \mathrm{curl}~\mathbf{v}$. These equations are part of a larger family of geophysical fluid models that can be derived from a variational principle and inherently reflect particular conservation laws \cite{holm2021stochasticmesoscale}. The governing equations (\ref{eq:Euler}) form a Lie-Poisson system \cite{marsden1983coadjoint} with a Hamiltonian $\mathcal{H}$ and an infinite number of conserved quantities, known as Casimirs $\mathcal{C}_k$, given by \begin{align}
        \mathcal{H}(\omega) &= -\frac{1}{2}\int \omega\psi, \\
        \mathcal{C}_k(\omega) &= \int \omega^k, \hspace{3mm} k=1,2,\ldots
\end{align}
where the integral is taken over the spatial domain.
\\
A discrete system with a similar Lie-Poisson structure is obtained after so-called geometric quantization. This structure-preserving discretization is based on a finite truncation of the Poisson bracket, as proposed in \cite{zeitlin1991finite, zeitlin2004self} and rests on the theory of quantization \cite{hoppe1989diffeomorphism, bordemann1994toeplitz, bordemann1991gl}. \sre{First, an $N>1$ is chosen, which can be thought of as the numerical resolution. Subsequently, a total of $\frac{N(N+1)}{2} - 1$ global basis functions are determined explicitly before carrying out a simulation. These functions serve to construct the discrete vorticity representation $W$. A} finite-dimensional approximation of the system \eqref{eq:Euler} is obtained as
\begin{equation}
\begin{aligned}
    \dot{W} &= [P, W],  \\
    \Delta_N P &= W.
    \end{aligned} \label{eq:QEuler}
\end{equation}
Here $W$ is the vorticity matrix, $P$ is the stream matrix and $W,P\in\mathfrak{su}(N)$, that is, skew-Hermitian, traceless $N\times N$ matrices. 

\sre{The discrete system \eqref{eq:QEuler} is interpreted as follows. A continuous vorticity field $\omega$ on the sphere can be expanded in a spherical harmonic basis $\{Y_{lm}\}$ as $\omega = \sum_{l, m} c_{lm} Y_{lm}$. The spherical harmonic coefficients $c_{lm}$ are used to construct the matrix $W$. Namely, \begin{equation}
    W = \sum_{l=0}^{N-1} \sum_{m=0}^l c_{lm}T_{lm}^N, \label{eq:quantized_expansion}
\end{equation} Here, $\{ T_{lm}^N \}$ is the so-called quantized spherical harmonic basis \cite{cifani2023efficient}, which provides a particular discrete approximation to the spherical harmonic basis $\{Y_{lm}\}$. In fact, the quantized representation enables the structure-preserving discretization \cite{zeitlin2004self, mclachlan1993explicit}. The basis element $T_{lm}^N$ is a sparse skew-Hermitian traceless matrix, nonzero only on the $m$-th sub- and superdiagonal. We refer to \cite{cifani2023efficient} for a detailed description of the quantized basis.} The quantized Laplacian $\Delta_N$ can be derived as a complicated expression, given in \cite{hoppe1998some}. \sre{The matrix $P$ then follows by applying the inverse quantized Laplacian to $W$.} The bracket $[P, W]= PW-WP$ is the standard matrix commutator.  In the limit of $N\rightarrow \infty$, the structure constants of the Lie algebra $\mathfrak{su}(N)$ converge to those of $C^\infty(\mathbb{S}^2)$ expressed in terms of spherical harmonics. \sre{This convergence implies that smooth functions on the sphere can be approximated by finite-dimensional matrices by means of Eq. \eqref{eq:quantized_expansion} \cite{cifani2023efficient}.} The discrete system is a Lie-Poisson system with a Hamiltonian $H$ and $N$ conserved quantities $C_k$,
\begin{align}
    H(W) &= \frac{1}{2}\mathrm{Tr}\left(PW\right), \\
    C_k(W) &= \mathrm{Tr}\left( W^k\right), \hspace{3mm} k=1,\ldots, N.
\end{align}
Equations \eqref{eq:QEuler} are solved numerically using the second-order isospectral midpoint rule \cite{viviani2019minimal, modin2020casimir}, using the parallelized implementation described in \cite{cifani2023efficient}. This is a Lie-Poisson integrator, conserving the $N$ discrete Casimir functions exactly. Given a time step size $h$, a time integration step proceeds as follows
\begin{equation}
\begin{aligned}
        W_n &= \left( I - \frac{h}{2} \Delta_N^{-1} \tilde{W} \right) \tilde{W} \left(I + \frac{h}{2}\Delta_N^{-1}\tilde{W} \right) \\
        W_{n+1} &= \left(I + \frac{h}{2}\Delta_N^{-1}\tilde{W}\right) \tilde{W} \left(I - \frac{h}{2}\Delta_N^{-1}\tilde{W} \right),
\end{aligned} \label{eq:ISOMP}
\end{equation}
i.e., given $W_n$ the intermediate solution $\tilde{W}$ is obtained first, after which $W_{n+1}$ is determined to complete a time-step.

\begin{figure}[h!]
    \centering
    \includegraphics[width=\textwidth]{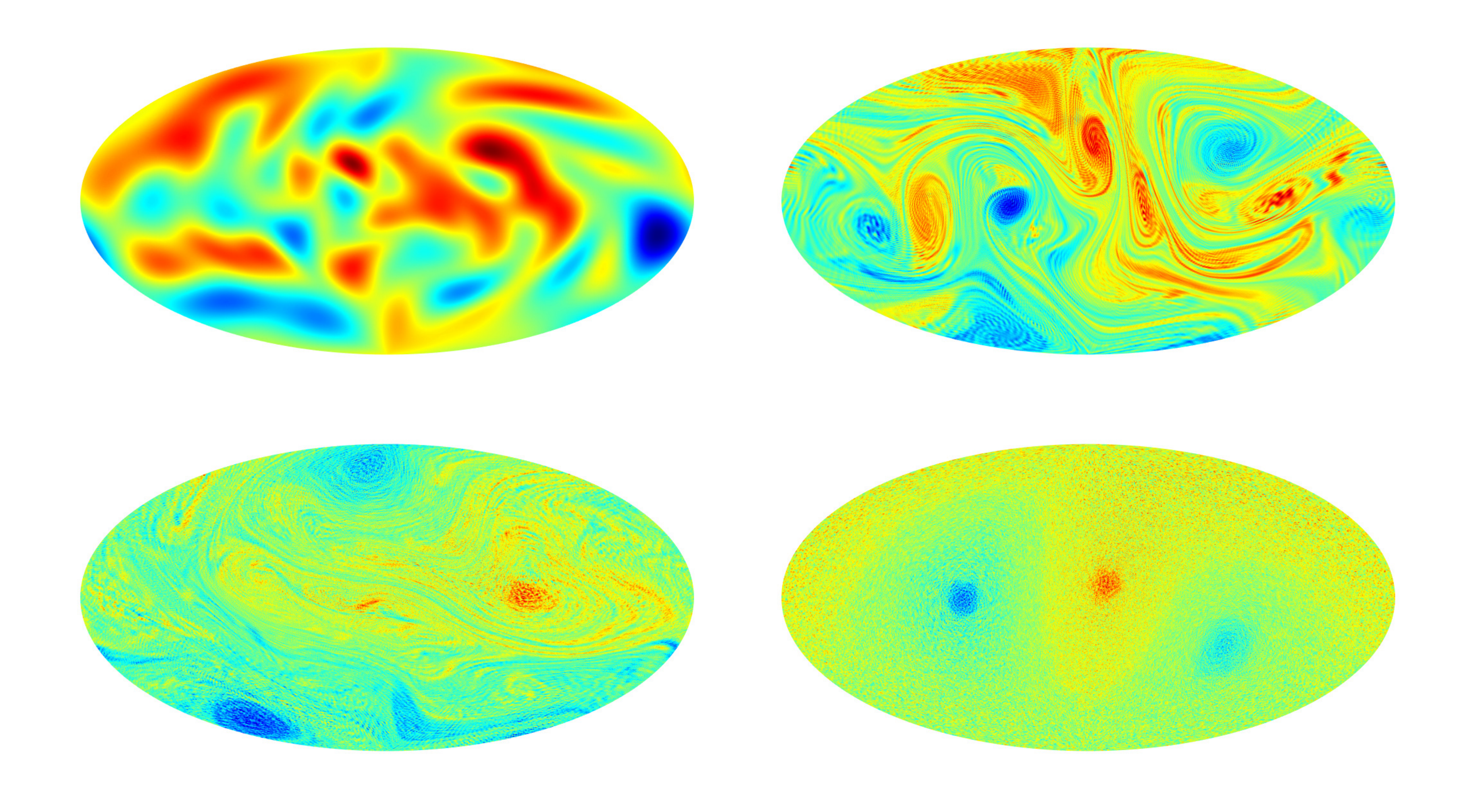}
    \caption{Snapshots of a high-resolution ($N=512$) numerical simulation of system \eqref{eq:QEuler}. The vorticity field is initialized as a random large-scale field (top left), after which it undergoes a period of vorticity mixing (top right, bottom left) before reaching a statistically steady state in which large-scale vorticity structures dominate the solution (bottom right).}
    \label{fig:fine_solution_spinup}
\end{figure}

An example of the Euler equations integrated at high resolution using \eqref{eq:ISOMP} is given in Fig. \ref{fig:fine_solution_spinup}. This figure shows the vorticity fields as Hammer projections in order to display the entire spherical domain. The flow dynamics reveal that large-scale low-dimensional structures are present in the vorticity field at late times \cite{modin2022canonical}. This motivates the use of coarse computational grids to capture the dynamics in the asymptotic time regime. A depiction of this is provided in Fig. \ref{fig:fine_solution_spinup}, showing late stages in the evolution of a high-resolution numerical simulation initialized from a random vorticity field in which only large scales are present. After a period of vorticity mixing the solution reaches a statistically steady state, in which large-scale vorticity structures have emerged and persist.

To compare numerical solutions at different resolutions, we define a fine-to-coarse filter. Throughout the paper the applied filter is a spectral cut-off filter, setting all coefficients corresponding to a wavenumber larger than a specified wavenumber to zero. In the following, we consistently choose a cut-off wavenumber defined by the coarse-grid resolution, which yields a filtered solution containing only spatial scales resolvable on the corresponding coarse grid.

\begin{figure}[h!]
    \centering
    \includegraphics[width=\textwidth]{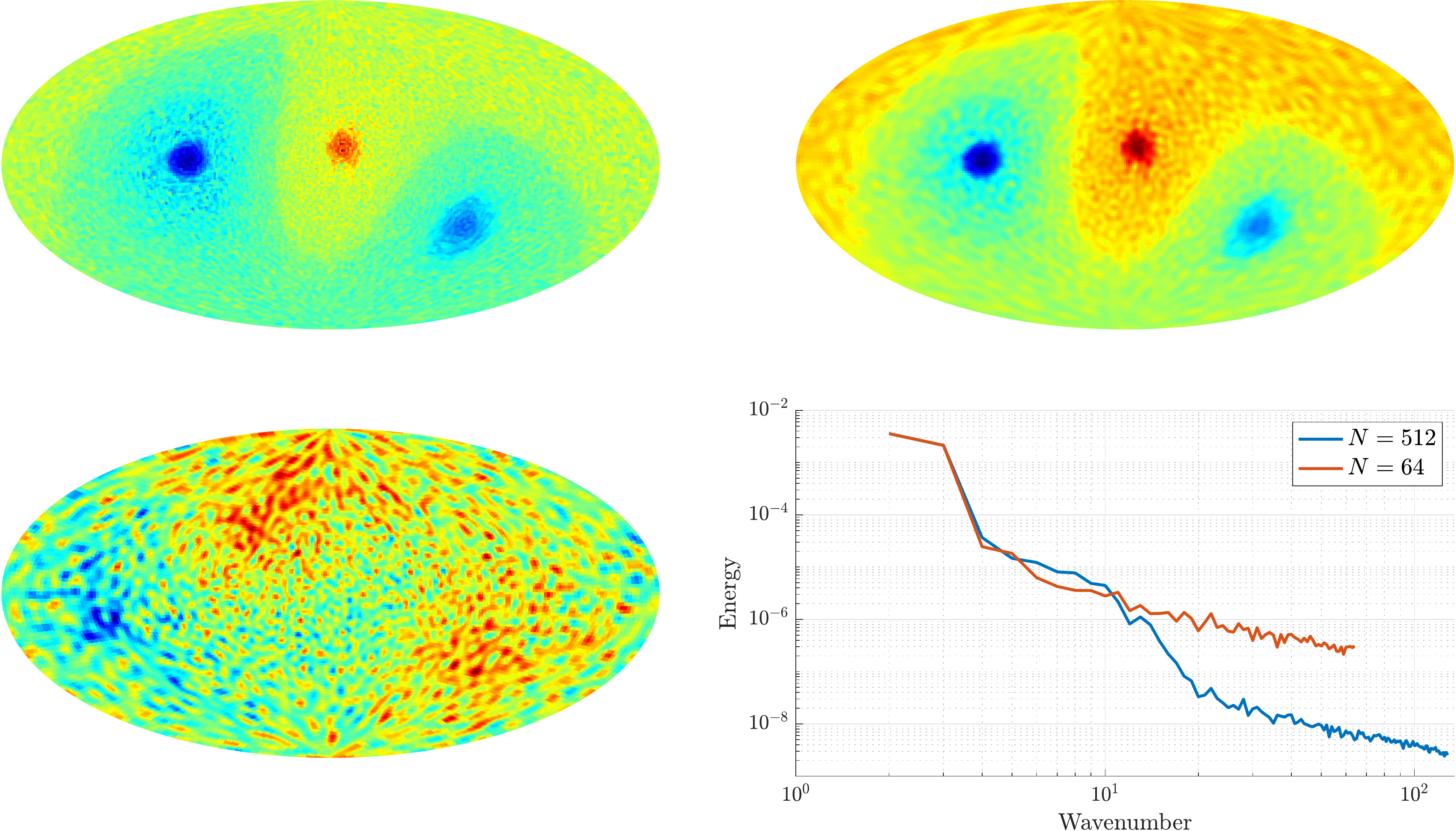}
    \caption{Snapshots of the fine vorticity field (top left), a filtered version thereof (top right) and a coarse vorticity field (bottom left) after reaching a statistically steady state. The energy spectra of the fine and coarse fields are shown in the bottom right panel.}
    \label{fig:W_Wr_Ws_spec_N128}
\end{figure}

Significantly decreasing the resolution yields a qualitatively different statistically steady state, as shown in Fig. \ref{fig:W_Wr_Ws_spec_N128}. Illustrated is a snapshot of the fine-grid solution ($N=512$), a filtered version thereof (only the components up to $N=64$ are shown), and a snapshot of a coarse-grid solution ($N=64$) using the algorithm as outlined in \eqref{eq:ISOMP}. These simulations are initialized using the smooth vorticity field in Fig. \ref{fig:fine_solution_spinup}. High-frequency components are visible in the snapshot of the high-resolution numerical solution, which develop as a result of the enstrophy cascade. By applying a spectral cut-off filter to the fine numerical solution we obtain a smooth vorticity field. By definition of the filter, this field can be fully resolved using the coarse resolution. Since the filtered fine solution is an orthogonal projection onto the coarse-resolvable subspace of solutions, it defines the best attainable result on the coarse grid and the result is a description of the large-scale components of the flow, influenced by all fine-grid resolvable scales. 

A clear qualitative difference exists between the filtered high-resolution vorticity and the vorticity obtained at a lower resolution, which is best explained by analyzing the energy spectra. The energy spectrum of the coarse numerical solution deviates from the spectrum of the fine numerical solution at the smallest resolvable scales. \sre{Nonetheless, the energy in the large scales is captured well. Additionally, the energy decay at large wavenumbers follows the same decay of $l^{-1}$ as observed in the high-resolution result, where $l$ is the wavenumber. Despite this agreement, the instantaneous vorticity field at the coarse resolution differs significantly from the high-resolution result.} The increased energy in high-frequency modes of the coarse numerical solution causes the small scales to dominate the vorticity field observed in Fig. \ref{fig:W_Wr_Ws_spec_N128}. We note that the energy spectrum of the filtered reference solution exactly coincides with the spectrum of the reference solution until the cut-off frequency at wavenumber 64, by definition of the filter. The discrepancy between the energy levels of the coarse numerical solution and the filtered reference may therefore be reduced by an appropriate forcing term. In the next section, we introduce a data-driven forcing term that yields the desired energy level at each frequency and thus regularizes the numerical solution.

\section{Data-driven spectrum-preserving forcing}\label{sec:model}
In the previous section, we observed that a qualitative difference exists between the statistical steady states obtained at low and high resolutions. One of the defining features of a statistically steady state is its kinetic energy spectrum. The corresponding energy spectra of the solutions reveal a considerable difference between the energy levels of the small scales present in the flow. The discrepancy between the spectra may be reduced by introducing an appropriate forcing or correction term to the coarse numerical simulations.  
In a statistically steady state, the spectrum is fully described using the mean and variance of the magnitude of the spectral coefficients in the statistically steady state. Therefore, the goal of the model is to reproduce these quantities accurately and, in doing so, recover the reference kinetic energy spectrum. In this section, we describe a forcing that achieves this goal, particularly in situations where the number of modes is kept low. For this purpose, we opt for a model that aims to match the mean and variance of the coefficient magnitudes to reference values. This approach is based on reference data corresponding to independently obtained highly resolved direct numerical simulations aiming to combine computational feasibility with accurate flow predictions. 

To define the spectral forcing, we expand the vorticity matrix in the quantized spherical harmonic basis $\left\{T_{lm}^N \right\}$,  \begin{equation}
    W(t) = \sum_{l=0}^{N-1} \sum_{m=0}^l c_{lm}(t) T_{lm}^N,
    \label{eq:spectral}
\end{equation}
with complex coefficients $\{c_{lm}\}$.
The energy in solution components at index $l$ is then defined as
\begin{equation}
    E_l(t) = \sum_{m=0}^l|c_{lm}(t)|^2.
    \label{eq:energy_wavenumber}
\end{equation}
The index $l$ will also be referred to as `wavenumber' following the analogy with an expansion in spherical harmonics and plane waves \cite{mehrem2011plane}.
The expansion \eqref{eq:spectral} allows us to formulate the equations of motion \eqref{eq:QEuler} in terms of the basis coefficients $c_{lm}$, as\begin{equation}
    \dot{c}_{lm} = L(\boldsymbol c, l, m),
\end{equation}
where $L$ is the spectral representation of $[P, W]$ and $\boldsymbol c$ the vector containing all basis coefficients. In particular, the evolution of the magnitude of $c_{lm}$ will be expressed as $L_r(\boldsymbol c, l, m)$. \sre{A special feature of our approach is that the time stepping acts on $W$, while the model is applied in spectral space. In the actual algorithm, a mapping between elements of $\mathfrak{su}(N)$ and their representation as quantized spherical harmonic coefficients is needed for this purpose. Therefore the operators $L$ and $L_r$ are not required to be explicitly defined or evaluated but serve only to simplify notation.} 

A mean-reverting forcing is introduced \sre{into the evolution of the coefficient magnitudes, to ensure that the magnitudes tend to a specified mean value. Forcing the magnitudes of the basis coefficients is pragmatic since these are stationary if the solution is in a statistically steady state. Mean reversion is realized} by adding an Ornstein-Uhlenbeck (OU) process to the evolution of the coefficient magnitude. This way, the reference spectrum can be reproduced in a coarse numerical simulation. It has been shown \cite{blomker2013accuracy} that the OU process arises in the governing equations as the continuous-time limit of  the 3D-var data assimilation algorithm \cite{courtier1998ecmwf}. We thus propose
\begin{equation}
    \text{d}|c_{lm}| = L_r(\boldsymbol c, l, m)\text{d}t + \frac{1}{\tau_{lm}}\left(\mu_{lm} - |c_{lm}| \right)\text{d}t + \sigma_{lm}\text{d}B_{lm}^t,
    \label{eq:spectral_forcing}
\end{equation}
where $\mu_{lm}$ and $\tau_{lm}$ are means and correlation times extracted from separate high-resolution simulation simulations. In fact, from a sequence of solution snapshots time series are obtained for each of the basis magnitudes $|c_{lm}|$, of which $\mu_{lm}$ is the mean value and $\tau_{lm}$ is the characteristic time scale. The relaxation of the forcing is determined by the time scale $\tau_{lm}$. Deviations of $|c_{lm}|$ from the mean $\mu_{lm}$ are nudged back in order to reduce the differences. Randomness is introduced via the term $\text{d}B_{lm}^t$ in which $B_{lm}^t$ is a general random process, defined for each pair $l, m$ separately. \sre{The random process can be tailored to fit the measurement data \cite{ephrati2023data}, though the common choice is to let $\text{d} B_{lm}^t$ be normally distributed with a variance depending on the time step size \cite{higham2001algorithmic}. We choose the latter in what follows and include the variance scaling in $\sigma_{lm}$.} The value of $\sigma_{lm}$ depends on the sample variance of the time series, on $\tau_{lm}$ and on the adopted time step size and will be specified later in this section. 


In the discrete setting, we apply the forcing defined by the OU process in \eqref{eq:spectral_forcing} as a correction after time step is completed. This alters a time-advancement step as follows. Starting from the vorticity $W^n$ at time level $t^n$, a prediction $\bar{W}^{n+1}$ of the vorticity at the next time level is obtained by integrating Eq. \eqref{eq:QEuler} over one time step using the algorithm \eqref{eq:ISOMP}. This prediction is then projected onto the basis of spherical harmonics to obtain the corresponding basis coefficients $\{\bar{c}^{n+1}_{lm}\}$. Finally, a correction is applied to these coefficients using \eqref{eq:spectral_forcing} to obtain $\{c_{lm}^{n+1}\}$ which are then used to construct the vorticity field $W^{n+1}$ at the new time level. We note that the correction is only applied to the magnitude of the basis coefficients. The parameter definitions in the implementation of \eqref{eq:spectral_forcing} will now be described.

The correction procedure \add{$(13)$} will be referred to as \textit{nudging}. We distinguish between \textit{deterministic nudging}, using only the deterministic component of the forcing, and \textit{stochastic nudging}, using both the deterministic and the stochastic component. The former is described as 
\begin{equation}
    |c_{lm}^{n+1}| = |\bar{c}_{lm}^{n+1}| + \frac{\Delta t}{\tau_{lm}}\left(\mu_{lm,\mathrm{det}} - |\bar{c}_{lm}^{n+1}|\right). \label{eq:deterministic_nudge}
\end{equation}
The stochastic nudge is defined as 
\begin{equation}
    |c_{lm}^{n+1}| = |\bar{c}_{lm}^{n+1}| + \frac{\Delta t}{\tau_{lm}}\left(\mu_{lm,\mathrm{stoch}} - |\bar{c}_{lm}^{n+1}|\right) + \sigma_{lm} \Delta B_{lm}^n, \label{eq:stochastic_nudge}
\end{equation}
where $\Delta B_{lm}^n$ is drawn from a standard normal distribution for each $l, m$ and $n$ independently.\\
The nudging procedures in Eqs. (\ref{eq:deterministic_nudge}, \ref{eq:stochastic_nudge}) can be characterized as a steady-state Kalman-Bucy filter \cite{grewal2014kalman} with prescribed gain $\Delta t / \tau_{lm}$. The value of $\tau_{lm}$ is chosen to be constant, similar to steady-state filters. At each time step, the `observation' consists of coefficients for each spherical harmonic mode separately. The deterministic nudging procedure assumes the observation is a fixed value $\mu_{lm, \mathrm{det}}$, whereas the stochastic approach adopts observations as distributed samples. Here, we use  $\mathcal{N}\left(\mu_{lm, \mathrm{stoch}}, \sigma_{lm}^2\right)$ as distribution and draw independent samples for each $l, m, n$ separately.
Thus, the unresolved interactions between different spherical harmonic modes are modeled as linear stochastic processes, independent for each value of $l, m$. This approach has been introduced as Fourier domain Kalman filtering \cite{majda2012filtering}. For low-dimensional systems it was analyzed in Fourier space \cite{harlim2008filtering, castronovo2008mathematical} and also shown to be feasible for filtering high-dimensional systems.\\
In the continuous formulation \eqref{eq:spectral_forcing} $\tau_{lm}$ can take on any positive value. In the discrete form (\ref{eq:deterministic_nudge}, \ref{eq:stochastic_nudge}), $\tau_{lm}$ can take on values in the interval $[\Delta t, \infty)$. For $\tau_{lm}=\Delta t$ the forcing ensures that the magnitude $|c_{lm}|$ of the corresponding coefficient becomes constant in the case of deterministic nudging. In the case of stochastic nudging, this value of $\tau_{lm}$ ensures that $|c_{lm}|$ evolves as Gaussian noise with the specified mean and variance. In the limit of large $\tau_{lm}$ the forcing approaches zero and the unforced dynamics is retained. \\
The nudging procedures in Eqs. (\ref{eq:deterministic_nudge}, \ref{eq:stochastic_nudge}) are treated as first-order autoregressive models with drift coefficient $(1-\Delta t/\tau_{lm})$ and mean $\mu_{lm,\mathrm{stoch}}$, which is a discretization of the OU process \eqref{eq:spectral_forcing}. 
The value of $\tau_{lm}$ is found by fitting the autocovariance function of the OU process to the sample autocovariance as obtained from the reference high-resolution simulation. The value of $\tau_{lm}$ is expected to decrease as larger wavenumbers $l$ are considered. This increases the contribution of the model term to the dynamics of the coefficients $c_{lm}$ at those wavenumbers. Therefore, with increasing spatial resolution one will resolve finer lengthscales associated with larger $l$, whose contributions correspond closer and closer to the direct observations. This is in accordance with theoretical results for filter performance \cite{majda2012filtering}.\\
The values of $\sigma_{lm},\mu_{lm,\mathrm{stoch}}$ and $\mu_{lm, \mathrm{det}}$ are chosen so that the reference energy spectrum is reproduced when the model is applied. Treating $|c_{lm}|$ as a stochastic variable, we observe that $\mathbb{E}(|c_{lm}|^2)$ is the expected energy content of the basis element $T_{lm}^N$. Through the definition of the variance we find that \begin{equation}
    \mathbb{E}(|c_{lm}|^2) = \mathrm{var}(|c_{lm}|) + \mathbb{E}(|c_{lm}|)^2
\end{equation}
We define $\sigma_{lm}$ so that variance of the autoregressive model coincides with the sample variance $s_{lm}^2$ of the reference time series, i.e., \begin{equation}
    \sigma_{lm} = s_{lm}\sqrt{1 - \left(1-\frac{\Delta t}{\tau_{lm}}\right)^2},
\end{equation}
where $s_{lm}$ is the sample standard deviation of $|c_{lm}|$ as obtained from the high-resolution simulation. To obtain the desired energy content, $\mu_{lm, \mathrm{stoch}}$ is subsequently chosen as $\mathbb{E}(|c_{lm}|)$. In the case of deterministic nudging, the variance of $|c_{lm}|$ vanishes when $\tau_{lm}=\Delta t$. To obtain the desired energy content in this limit, $\mu_{lm,\mathrm{det}}$ is chosen as $\sqrt{\mathbb{E}(|c_{lm}|^2)}$. The mean, variance, and correlation time are estimated using standard unbiased estimators of which the mean squared error decreases linearly with the number of used samples. It is assumed that the mean and variance are constant in time, therefore requiring that the flow is in a statistically steady state.

For each basis function only three parameters need to be measured: the mean, the variance and the correlation time. This outlines the simplicity of the model. These parameters are inferred from the data, do not require additional tuning and are defined up to the resolution of the reference solution. Furthermore, the basis of spherical harmonics is resolution-independent. Therefore the forcing parameters only depend on the reference data and not on the choice of the coarse-grid resolution or time step size, implying that the model is self-consistent \cite{frederiksen2017}. This is further corroborated in a later section of the paper by applying the model at various low resolutions. 

\section{Numerical experiments}\label{sec:results}
In this section, we apply the forcing proposed in Section \ref{sec:model} to coarse numerical simulations. We describe the reference solution and introduce the measured variables that constitute the model data. The forcing is applied at different coarse computational grids using several model configurations. The model results are compared to the reference solution and the no-model coarse numerical solution and are assessed in terms of statistical quantities of the resulting time series of the basis coefficients. Finally, we illustrate that application of the model yields accurate long-time solutions on coarse computational grids.

\subsection{Description of reference solution} \label{eq:reference}
\begin{figure}[h!]
    \centering
    \includegraphics[width=\textwidth]{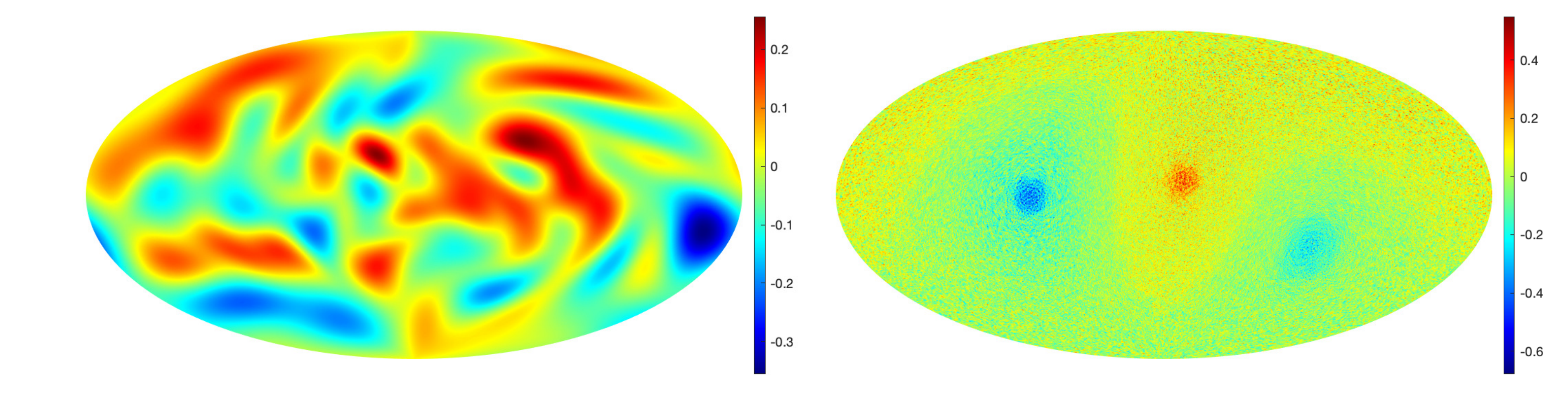}
    \caption{Left: Initial vorticity field used in the numerical simulations performed throughout the paper. Right: Snapshot of the vorticity field after reaching a statistically steady state.}
    \label{fig:IC}
\end{figure}

\begin{figure}[h!]
    \centering
    \includegraphics[width=\textwidth]{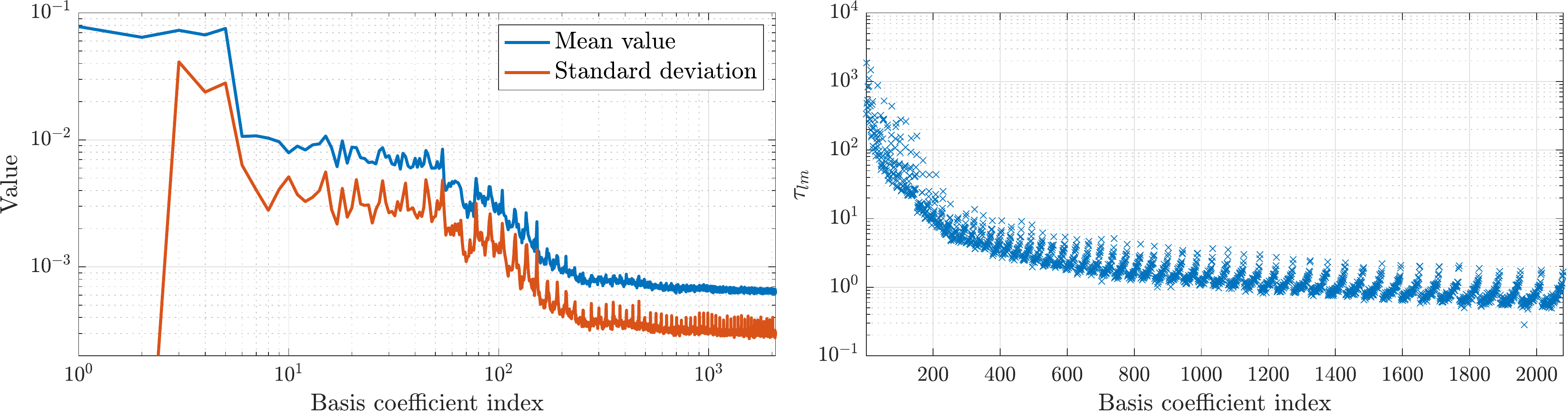}
    \caption{Left: measured means and standard deviations of the absolute value of each basis coefficient of the reference solution. Right: Estimated correlation time for each basis coefficient of the reference solution.}
    \label{fig:measured_variables}
\end{figure}

The reference solution is acquired from the discretized equations described in Section \ref{sec:governingeqs} and adopts a resolution $N=512$. The initial condition is the smooth vorticity field as shown in the left panel of Fig. \ref{fig:IC}, which is also adopted in later numerical simulations using lower resolutions. This initial condition is randomly generated and contains only large scales of motion. The vorticity is evolved until $t=6500$, shown on the right panel of Fig. \ref{fig:IC}, at which a statistically steady state is reached. This was verified by averaging the kinetic energy spectrum over several time durations. High-resolution snapshots are collected every time unit after reaching this state. A total of 1000 snapshots is collected to ensure that estimates of the mean, variance, and correlation times are sufficiently accurate. \\
By projecting the snapshots onto the basis $\{T_{lm}^N \}$ a time series of coefficients for each spherical harmonic mode is obtained. These coefficients are complex-valued, however, in what follows we will only consider the time series of the corresponding magnitudes since these are the quantities that the proposed model acts on.\\
The forcing parameters are shown in Fig. \ref{fig:measured_variables}, sorted per basis coefficient. Here, we show the measured means, standard deviations, and correlation times that are used in the model. On a grid of resolution $N$ a total of $N(N+1)/2 - 1$ basis functions $T_{lm}^N$ is available, which can be sorted in ascending order of $l$ and $m$. Here only the first 2079 values are shown, corresponding to all resolvable modes for $N=64$, a coarse resolution that will be investigated momentarily. A decreasing mean value and variance are observed as the scale size is decreased. This is seen until basis functions with $l=23$ are considered, at the $275^\mathrm{th}$ basis coefficient, after which the mean and variance remain roughly constant. This corresponds to the wavenumber at which the reference energy spectrum follows the $l^{-1}$ decay. The measured correlation time $\tau_{lm}$ becomes smaller as larger wavenumbers are considered, which indicates that the smaller scales in the flow behave in an increasingly dynamic manner. The small values of $\tau_{lm}$ result in a relatively larger contribution from the model term to the dynamics of the smallest resolvable scales.

\subsection{Coarse-grid flow simulations}
In this subsection, the performance of the model is tested in coarse-grid numerical simulations of the flow. In particular, the model is applied at resolutions $N=64$ and $N=32$ to show that the forcing parameters are applicable at different coarse resolutions. The chosen levels of coarsening provide a significant reduction in computational costs. At the same time, the dominant flow patterns can be accurately resolved, as shown in Section \ref{sec:governingeqs}. Four different settings for the model are studied by varying the minimal wavenumber at which the model is activated and by either enabling or disabling the stochastic model term. The scales at which the model is applied are $l\geq 1$ and $l\geq 8$ for resolutions $N=64$ and $N=32$, in order to capture the same flow complexity at different resolutions. The choice of $l\geq 1$ corresponds to applying the model at all available scales, whereas $l\geq 8$ only applies to small-scale flow features. For each resolution, we illustrate the need for modeling by providing snapshots of the filtered reference solution and the no-model coarse-grid solution. From these figures, the qualitative features of the solution at different resolutions become apparent.

\begin{figure}[h!]
    \centering
    \includegraphics[width=\textwidth]{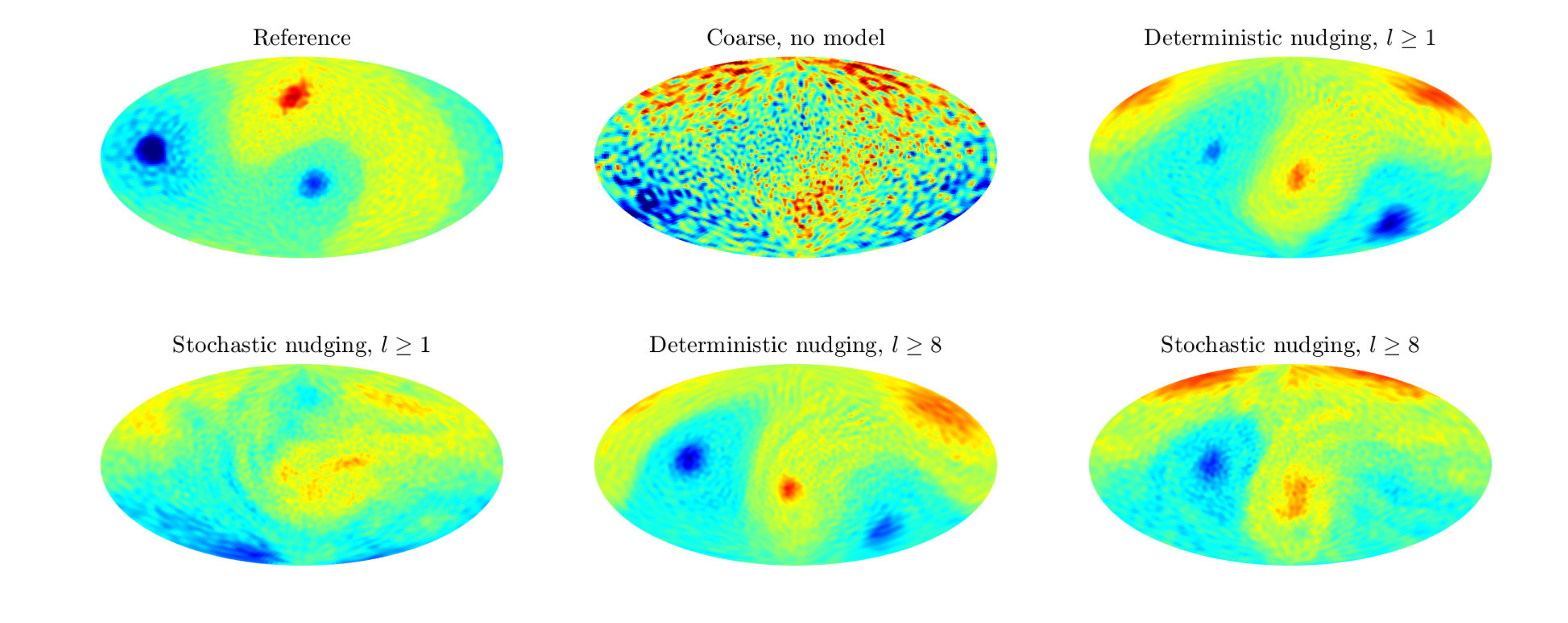}
    \caption{Snapshots of numerical solutions at a statistically steady state. Top left: filtered reference solution, displaying only modes resolvable for $N=64$. Top middle: no-model coarse numerical solution. Top right and bottom row: coarse numerical solution with forcing applied, using the full model term or only the deterministic part, with varying minimal wavenumber at which the forcing is applied.}
    \label{fig:snapshots_model_N64}
\end{figure}

\begin{figure}[h!]
    \centering
    \includegraphics[width=0.48\textwidth]{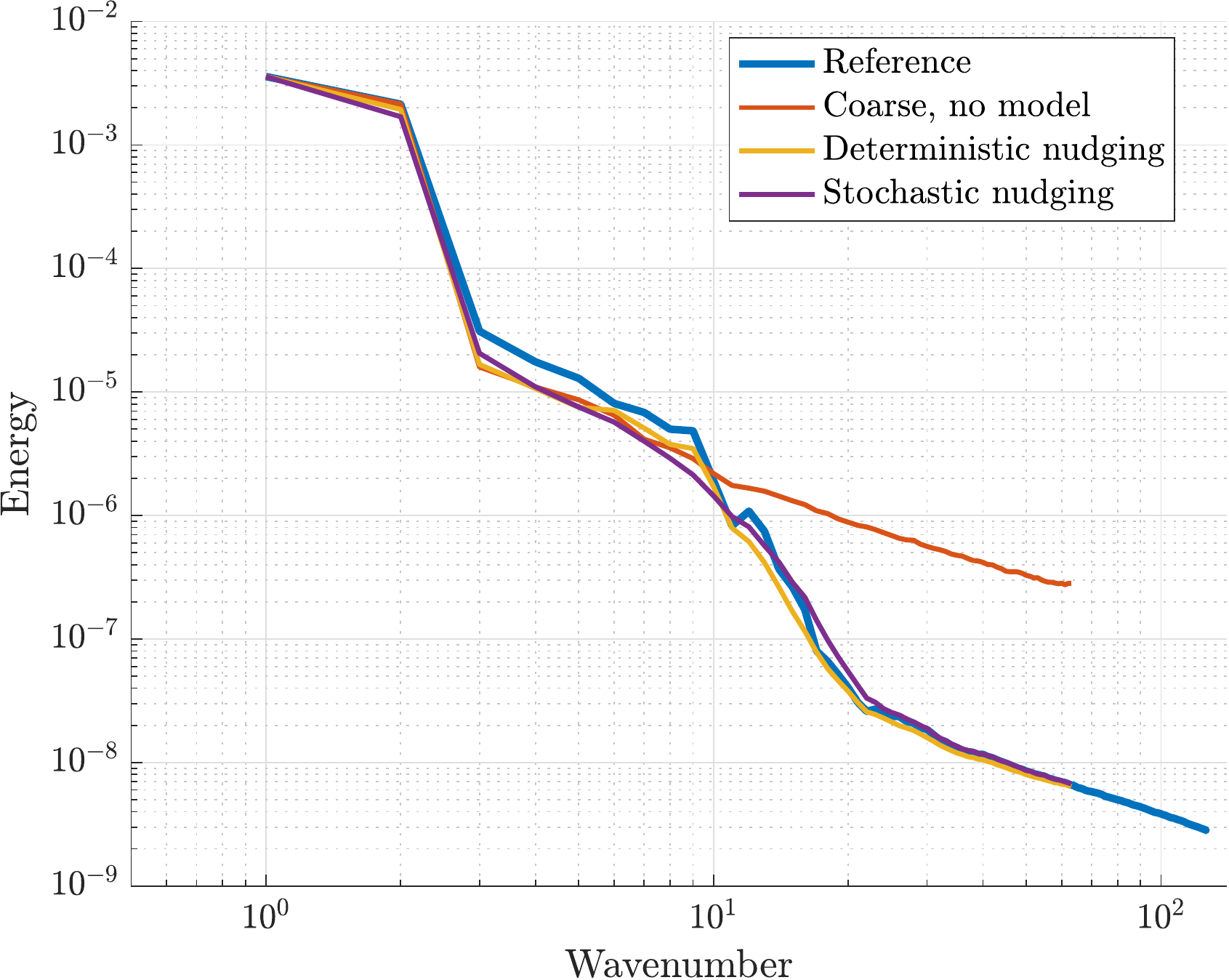}
    \includegraphics[width=0.48\textwidth]{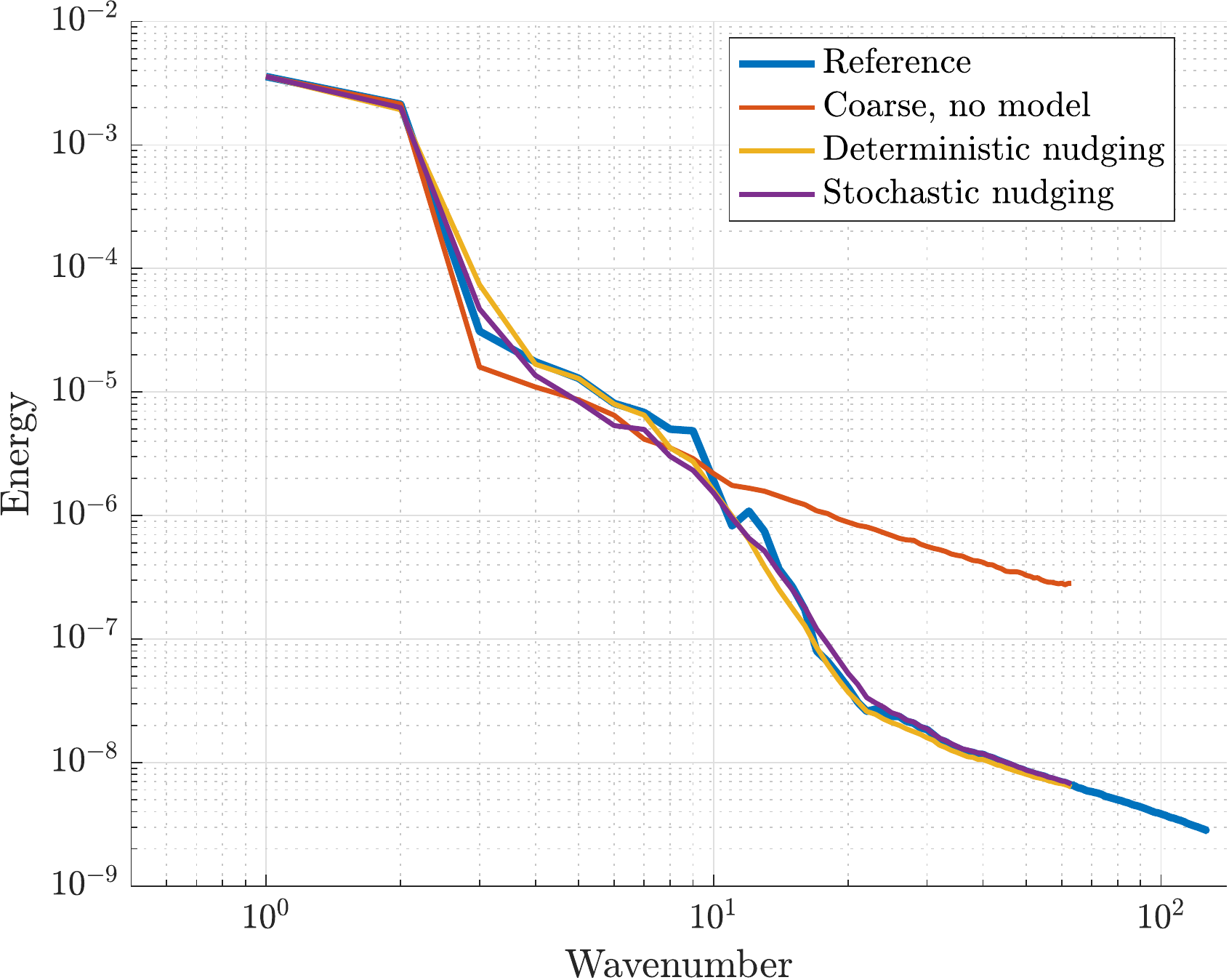}
    \caption{Average energy spectra for forced coarse solutions, using $N=64$, compared to the energy spectra of the reference solution and the no-model coarse solution. The forcing is applied at wavenumbers $l\geq 1$ (left) and wavenumbers $l\geq 8$ (right).}
    \label{fig:energy_spectra_model64}
\end{figure}

\begin{figure}[h!]
    \centering
    \includegraphics[width=\textwidth]{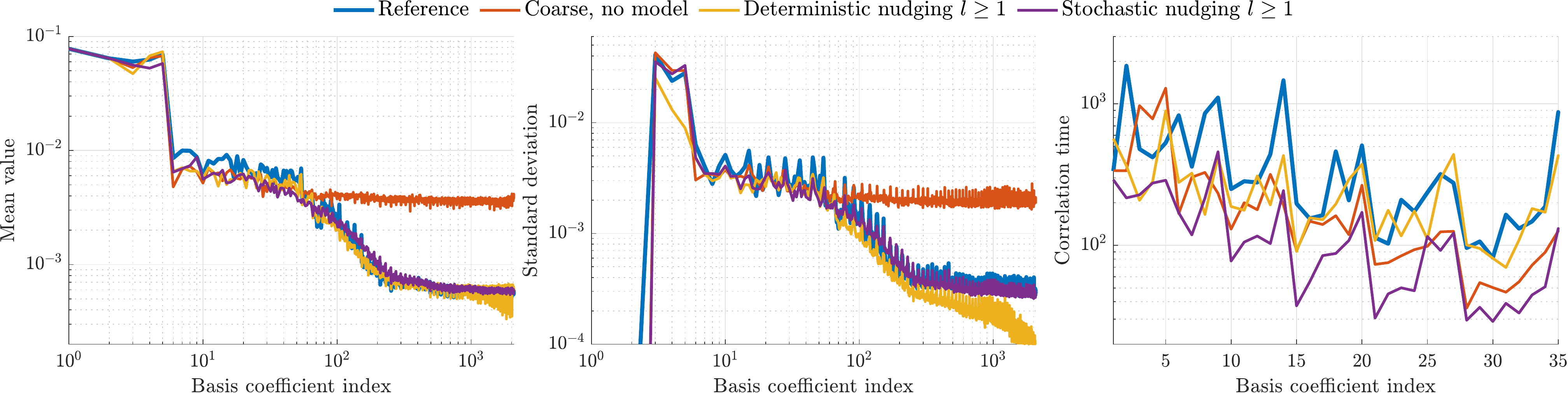}
    \centering
    \vspace{5mm}
    \includegraphics[width=\textwidth]{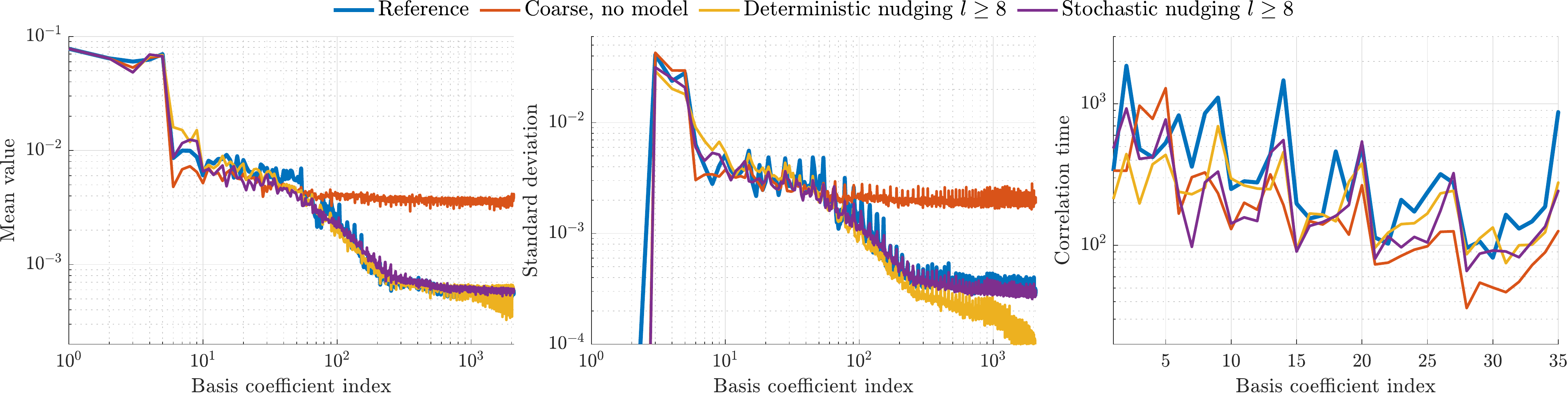}
    \caption{Statistics of the basis coefficient time series of the reference solution, no-model coarse solution, and coarse solution with the model applied for $N=64$. Shown here are the results when applying the model at wavenumbers $l\geq 1$ (top row) and $l\geq 8$ (bottom row). The mean value (left) and standard deviation (middle) are shown for all wavenumbers. The correlation time (right) is shown for the large-scale components, with wavenumbers $l\leq 8$.}
    \label{fig:model_statistics_64}
\end{figure}

We first consider the results at resolution $N=64$. A qualitative comparison of the different numerical solutions is provided in Fig. \ref{fig:snapshots_model_N64}. The top left panel shows a snapshot of the reference solution at the statistically steady state, where the high-frequency components have been filtered from the solution. As before, the applied filter is a spectral cut-off filter where the cut-off wavenumber is defined by the coarse-grid resolution. Coherent large-scale vorticity structures that are resolvable on the coarse grid are visible in this snapshot. The merit of the forcing can be observed through the differences between the coarse numerical simulation results. The top middle panel shows the no-model coarse solution, which shows clear qualitative differences with the reference result. The top right panel and the bottom row show forced coarse numerical solutions at statistically steady states, using the different forcing settings. Evidently, the latter snapshots reveal a smoother vorticity field and a more accurate representation of the reference vorticity, compared to the coarse no-model simulation. In particular, a qualitative agreement in terms of large-scale vortex structures may be observed. In this specific case, a large connected positive vorticity structure (in red) and two smaller negative vorticity structures (in blue) are reproduced when applying the model. Interestingly, the proposed nudging concentrates some additional positive vorticity in the tail of the coherent structure (in red), whereas no such behavior is observed for the negative vorticity.
\\
The qualitative differences are reflected in the energy spectra, visualized in Fig. \ref{fig:energy_spectra_model64}, showing the energy spectra using the forcing for $l\geq 1$ and $l\geq 8$ in the two panels. By construction, nudging reduces the energy content in the small scales of the flow. Accurate energy levels are observed for both the deterministic and stochastic nudging procedures. These results are observed for both choices of scales at which the forcing is applied. A good agreement in the energy at the large scales is observed for all performed simulations. Particularly, the energy spectra demonstrate a striking agreement at the smallest resolved scales when the model is applied. This suggests that the choice of parameters for the deterministic and stochastic forcing is well-suited for reproducing the energy spectra at these scales.
\\
A quantitative comparison of the statistics of the solutions is given in Figure \ref{fig:model_statistics_64}. For each basis coefficient, the mean, standard deviation and estimated correlation time are shown. The mean and the standard deviations of the time series display similar qualitative behavior regardless of the minimal wavenumber at which the forcing is applied. For these quantities, both the deterministic nudging and the stochastic nudging lead to significant improvement compared to the no-model results. Including the stochastic component of the forcing, based on the high-resolution reference data, leads to an increased agreement at the smaller scales of the flow, indicating that the inclusion of additional variance in the forcing of the small scales leads to a truthful reproduction of these statistical quantities.

The estimated correlation times of the large-scale modes $(l\leq 8)$ in Fig. \ref{fig:model_statistics_64} show that deterministic nudging of all modes yields an improved correlation time compared to the no-model case. However, the stochastic nudging procedure for $l\geq 1$ leads to smaller correlation times compared to the coarse no-model simulation, implying that the stochastic component of the forcing is too strong. A qualitative improvement is observed when applying the model to wavenumbers $l\geq 8$, for both deterministic and stochastic nudging. These results suggest that the evolution of large scales in the flow benefits from an accurate statistical description of the evolution of small scales. This coincides with a basic premise underlying large-eddy simulation. 

The numerical experiment is repeated at a resolution $N=32$ to demonstrate that the proposed model yields forcing parameters that can be efficiently applied at different coarse resolutions. At this resolution, large spatial structures in the flow may still be resolved with acceptable accuracy. The forcing will be applied at wavenumbers $l\geq 1$ and $l\geq 8$, where the model affects all scales of motion in the former and only the small scales in the latter. 

\begin{figure}[h!]
    \centering
    \includegraphics[width=\textwidth]{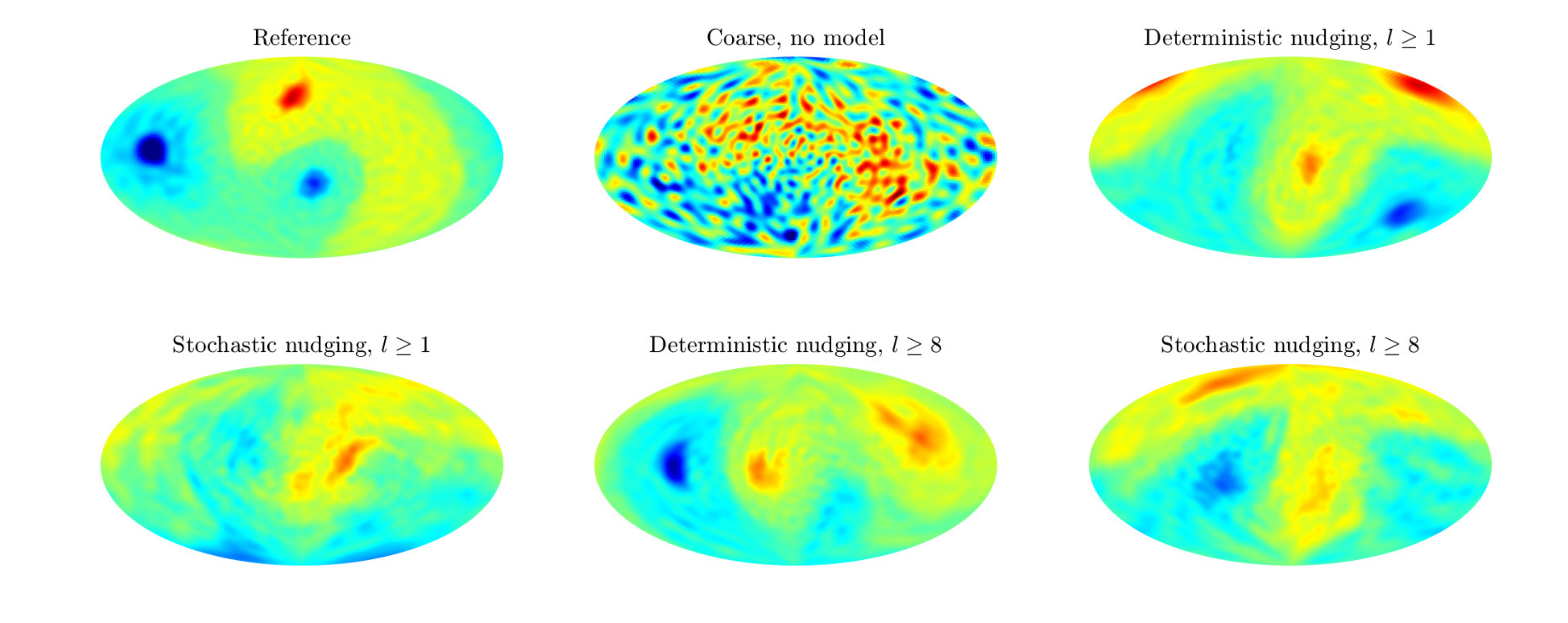}
    \caption{Snapshots of numerical solutions at a statistically steady state. Top left: filtered reference solution, displaying only modes resolvable for $N=32$. Top middle: no-model coarse numerical solution. Top right and bottom row: coarse numerical solution with forcing applied, using the full model term or only the deterministic part, with varying minimal wavenumber at which the forcing is applied.}
    \label{fig:snapshots_model_N32}
\end{figure}

\begin{figure}[h!]
    \centering
    \includegraphics[width=0.48\textwidth]{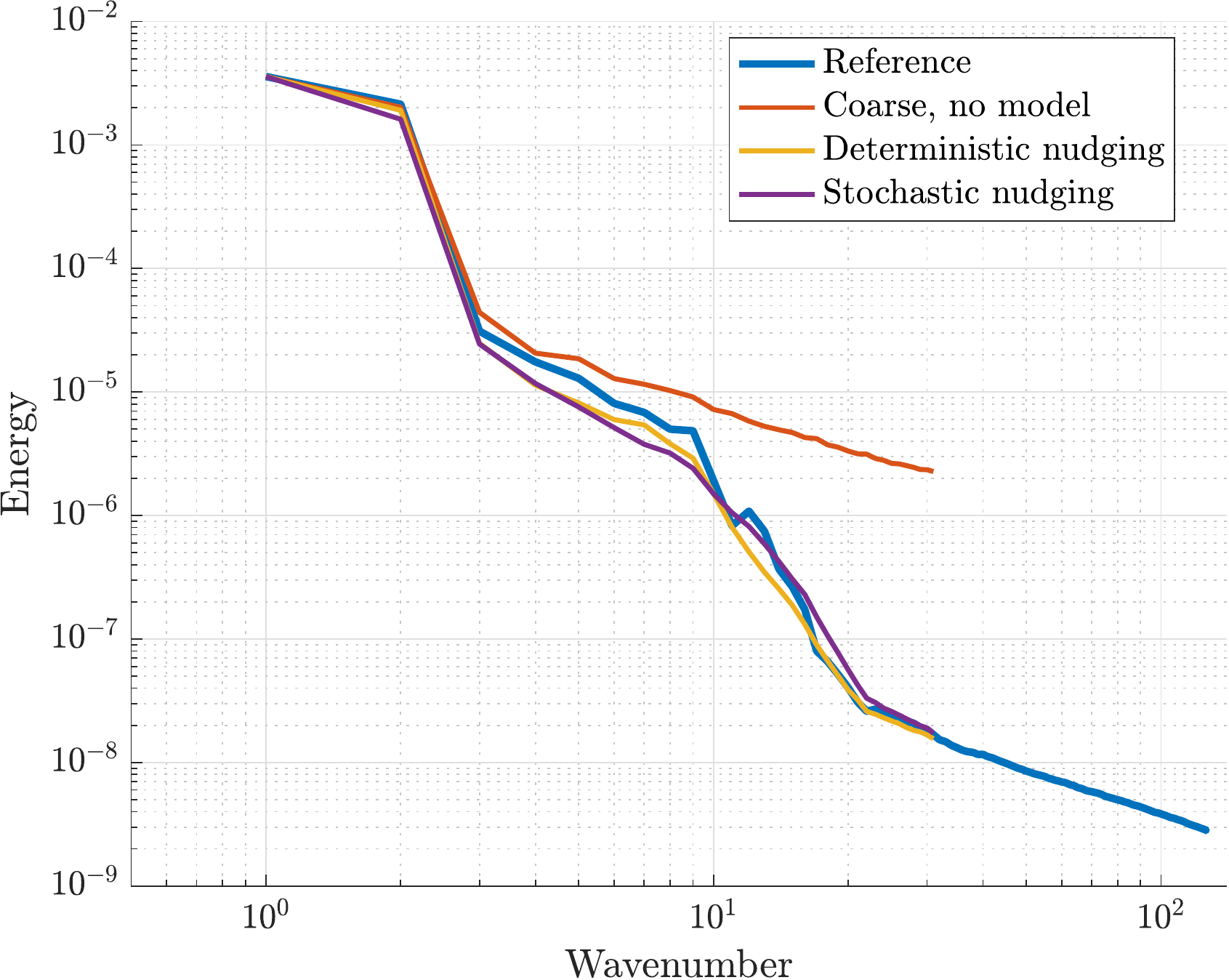}
    \includegraphics[width=0.48\textwidth]{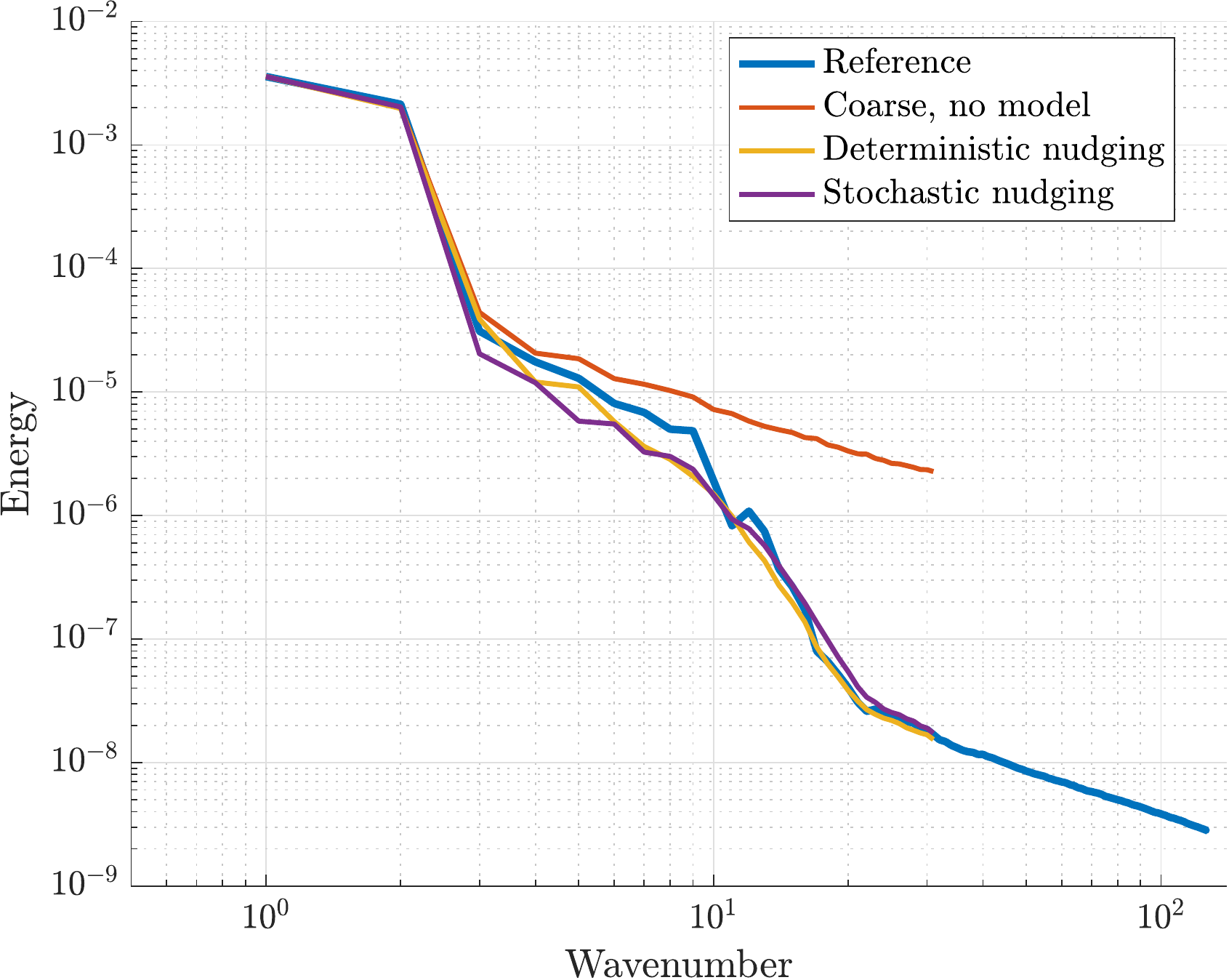}
    \caption{Average energy spectra for forced coarse solutions, using $N=32$, compared to the energy spectra of the reference solution and the no-model coarse solution. The forcing is applied at wavenumbers $l\geq 1$ (left) and wavenumbers $l\geq 8$ (right).}
    \label{fig:energy_spectra_model32}
\end{figure}

A qualitative comparison of the statistically steady states is given in Fig. \ref{fig:snapshots_model_N32}. It may be seen that the model effectively produces a smooth vorticity field with qualitatively similar features as the reference solution. This is reflected by the decrease of energy in the smallest resolvable scales compared to the no-model formulation, as shown in the energy spectra in Fig. \ref{fig:energy_spectra_model32}. As previously observed, all coarse-grid numerical simulations accurately capture the energy in the largest scales of motion. Applying the model also leads to a notable agreement with the reference solution in the average energy levels of the smallest resolvable scales.  

\begin{figure}[h!]
    \centering
    \includegraphics[width=\textwidth]{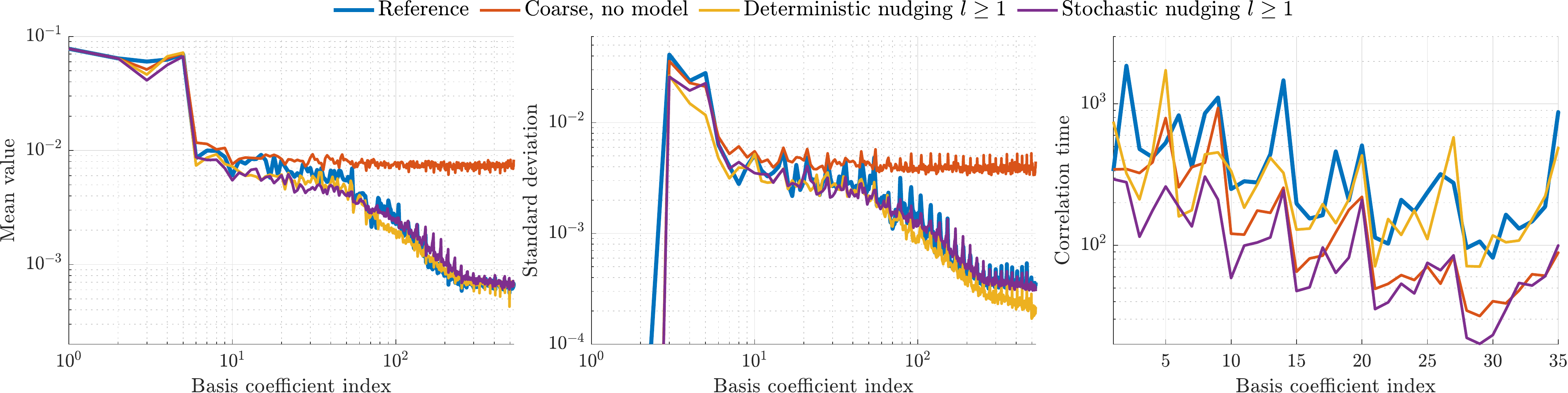}
    \centering
    \includegraphics[width=\textwidth]{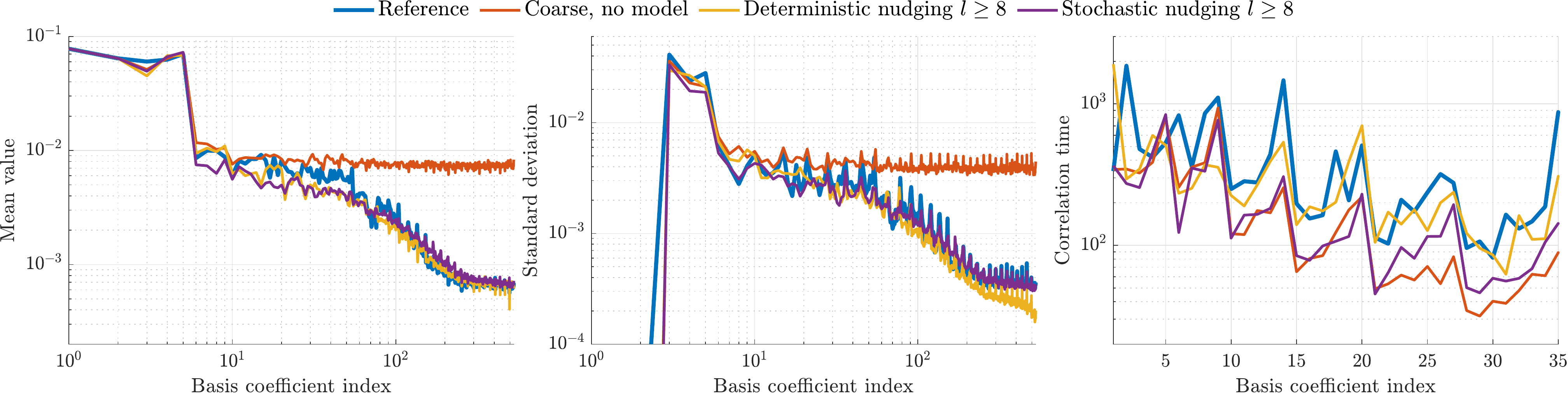}
    \caption{Statistics of the basis coefficient time series of the reference solution, no-model coarse solution, and coarse solution with the model applied for $N=32$. Shown here are the results when applying the model at wavenumbers $l\geq 1$ (top row) and $l\geq 8$ (bottom row). The mean value (left) and standard deviation (middle) are shown for all wavenumbers. The correlation time (right) is shown for the large-scale components, with wavenumbers $l\leq 8$.}
    \label{fig:model_statistics_32}
\end{figure}

A comparison of the mean value, standard deviation and estimated correlation times of the time series of the basis coefficients is given in Fig. \ref{fig:model_statistics_32}. Applying the model leads to a clear improvement in the mean and variance of the coefficients, regardless of the choice of length scales at which the forcing is applied. Employing the deterministic forcing at all lengthscales yields a good agreement in the correlation times, whereas the stochastic forcing reduces the measured correlation times and yields no improvement. The correlation times are also found to improve when applying the deterministic model only to components with wavenumber $l\geq 8$. The stochastic forcing displays no significant improvement when applied at these wavenumbers.

\subsection{Large-scale vortex dynamics at statistically steady states}
\begin{figure}[h!]
    \centering
    \includegraphics[width=\textwidth]{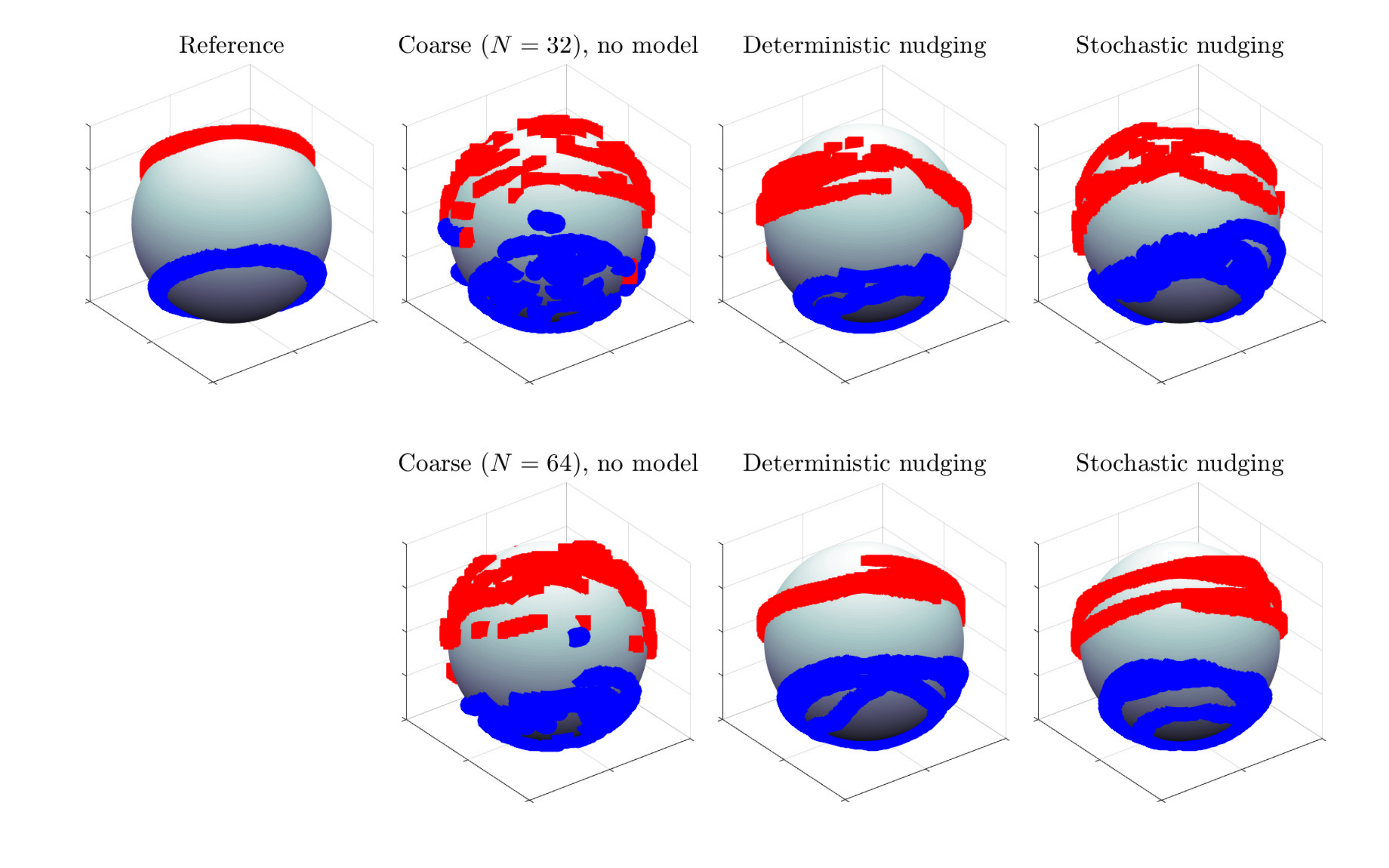}
    \caption{Trajectories of large-scale vortices for various numerical solutions. The red and blue lines denote the trajectories of the maximum and minimum vorticity values, respectively. Shown are the measured trajectories of the reference solution (top left). The coarse no-model numerical solution and the coarse solution with the model at resolution $N=32$ are displayed on the top row. Both the deterministic and the stochastic models are applied to wavenumbers $l\geq 8$ at this resolution. The bottom row shows the realizations at resolution $N=64$, where the model is applied to wavenumbers $l\geq 8$. }
    \label{fig:blob_trajectories}
\end{figure}

The qualitative predictions of coarse-grained modeled dynamics can be analyzed by means of the vortex trajectories over long integration times. Here, the vortex movement is tracked by locating the maximum and minimum attained vorticity value at each solution snapshot. According to \cite{modin2020casimir}, high-resolution numerical experiments indicate that the ratio between the angular momentum and enstrophy governs the number of large-scale vortex structures in the final statistically steady state. This ratio is determined by the initial condition and remains constant throughout the no-model numerical simulations since the angular momentum and the enstrophy are conserved quantities in the discretized system. Additionally, the vortex trajectories are found to be stable. Thus, the long-term qualitative behavior of the coarse numerical solutions can be assessed by measuring the number of large-scale vortices and their trajectories. As we previously observed, the coarse-grained modeled vorticity fields show qualitative agreement in terms of the number of vortices. Here, we demonstrate the capability of the model to accurately yield stable long-time vortex dynamics by tracking vortex movement over long simulation times. \\
The long-time vortex trajectories for various numerical realizations are shown in Fig. \ref{fig:blob_trajectories}. The reference trajectories are obtained from the high-resolution simulation as used in the previous section. The model results at resolutions $N=32$ and $N=64$ are obtained by applying the model to wavenumbers $l\geq 8$. The reference trajectories display stable movement along clearly defined trajectories about a fixed axis. Such behavior is not observed for the coarse no-model results, where instead the extreme values of the vorticity move in a seemingly unorganized fashion without distinct trajectories. Applying the model to either of the presented resolutions yields a noticeable qualitative improvement in the measured vortex movement. In particular, we identify trajectories about the same fixed axis as the reference trajectories but the model trajectories exhibit perturbations. The perturbations appear stronger when stochastic nudging is applied and when coarser grids are considered.

\section{Conclusions and outlook} \label{sec:conclusion}
In this paper, we have proposed and assessed a standalone data-driven model for the coarsening of the Euler equations on the sphere. High-resolution simulation snapshots were used as a reference. This data was decomposed into spherical harmonic modes and corresponding time series of coefficients were determined. A stochastic model was introduced to compensate for shortcomings introduced by severe coarsening. The model parameters were obtained from statistics of the spherical harmonic coefficients time series. In particular, the proposed model was designed to reproduce the kinetic energy spectrum of the reference data in statistically steady states by adopting a nudging strategy similar to continuous data assimilation. \\
The model is imposed using a prediction-correction scheme leading to a formulation similar to a steady-state Fourier domain Kalman filter. We opted for a separate nudging strength for each of the forced lengthscales, dependent on the corresponding measured characteristic timescale, and demonstrated that this approach accurately recovers the energy levels in small resolved spatial scales and leads to stable long-time solutions. Moreover, no assumptions about the employed resolution are used in the derivation model. This was demonstrated by first measuring the forcing parameters and subsequently applying the model on several coarse computational grids. The proposed stochastic and deterministic models were not found to differ much in terms of results. Both approaches yielded accurate kinetic energy spectra at strong coarsening. In addition, the deterministic model yielded accurate correlation times of the magnitudes of the spherical harmonic basis coefficients, indicating accurate evolution of the large-scale flow features. \\ 
The results in this paper show that the decomposition of a high-resolution reference signal into spatial global basis functions and temporal coefficients can be employed efficiently to obtain resolution-independent forcing parameters to be used in models for coarse numerical simulations. The proposed model relies on several simplifying assumptions which will be scrutinized in future work. In particular, the robustness of the model in terms of stability and accuracy with respect to varying nudging strengths will be assessed. The connection to data assimilation algorithms and Kalman filtering theory may help to extend the model and weaken underlying assumptions, for example by including estimated covariance between different spherical harmonic modes in the model. \\
The approach presented here is general for flows in statistically stationary states and is not restricted to the two-dimensional Euler equations or the use of spherical harmonic modes as a global basis. Different flow settings may be considered by adopting, for example, a Fourier basis for periodic domains or, more generally, proper orthogonal decomposition (POD) modes when boundaries are present in the domain. Further work will be dedicated to extending the proposed model to different sophisticated flow settings such as two-dimensional Rayleigh-B\'enard convection, the rotating Euler equations on the sphere, or the quasi-geostrophic equations on the sphere.

\subsection*{Acknowledgements}
The authors would like to thank Darryl Holm, at the Department of Mathematics, Imperial College London, Erwin Luesink and Arnout Franken, at the University of Twente, and Klas Modin, at Chalmers University of Technology, for their valuable input and discussions in the context of the SPRESTO project, funded by the Dutch Science Foundation (NWO) in their TOP1 program.

\bibliographystyle{plain}
\bibliography{refs_paper}

\end{document}